\documentclass[a4paper]{article}

\usepackage{INTERSPEECH2020}

\title{Understanding effect of speech perception in EEG based speech recognition systems}
\name{Gautam Krishna, Co Tran, Mason Carnahan, Ahmed H Tewfik}
\address{
  Brain Machine Interface Lab, The University of Texas at Austin
  }
\email{}

\begin{document}

\maketitle
\begin{abstract}
The electroencephalography (EEG) signals recorded in parallel with speech are used to perform isolated and continuous speech recognition. During speaking process, one also hears his or her own speech and this speech perception is also reflected in the recorded EEG signals. In this paper we investigate whether it is possible to separate out this speech perception component from EEG signals in order to design more robust EEG based speech recognition systems. 
We further demonstrate predicting EEG signals recorded in parallel with speaking from EEG signals recorded in parallel with passive listening and vice versa with very low normalized root mean squared error (RMSE). We finally demonstrate both isolated and continuous speech recognition using EEG signals recorded in parallel with listening, speaking and improve the previous connectionist temporal classification (CTC) model results demonstrated by authors in \cite{krishna2019state} using their data set.   

\end{abstract}

\section{Introduction}

The electroencephalography (EEG) signals are non invasive neural signals which are recorded by placing EEG sensors on the scalp of the subject. The EEG signals demonstrate high temporal resolution even though the spatial resolution and signal to noise ratio (SNR) demonstrated are poor. Since EEG is a safe non invasive technique it is easy to test and deploy EEG based brain computer interface (BCI) systems. In the recent years there has been lot of interest in research community in trying to develop speech recognition systems using EEG signals where EEG signals recorded in parallel with speech are translated to text using automatic speech recognition (ASR) models. For example the work demonstrated by authors in \cite{krishna2019speech} demonstrates isolated speech recognition using EEG signals on a limited English vocabulary consisting of four words and five vowels. Similarly in \cite{krishna20} authors demonstrate continuous speech recognition using EEG signals on a limited English vocabulary consisting of 20 unique sentences. In a more recent work explained in \cite{krishna2020synthesis,krishna2019state} authors demonstrated preliminary results for synthesizing speech directly from EEG features. In \cite{krishna2019state} authors introduced new EEG feature sets and also demonstrated continuous speech recognition using EEG signals recorded in parallel with speech and listening on a limited English vocabulary consisting of 9 unique sentences. 
The potential benefits of EEG based speech recognition systems include overcoming performance loss of ASR systems operating in presence of background noise \cite{krishna2019speech}, helping with improving technology accessibility for people with speaking disabilities or people who are not able to produce voice by allowing them to use voice assistant systems trained to recognize EEG or combination of EEG and acoustic features.  

Even though in references \cite{krishna20,krishna2019state,krishna2019improving,krishna2019speech} authors used EEG signals recorded in parallel with speaking for performing speech recognition, during speaking process a person also gets feedback via listening or hearing his or her own speech. Thus the EEG signals recorded in parallel with speaking includes the brain activity responsible for speech production as well as speech perception. In this paper we propose a deep learning model to separate the perception component from the EEG signals recorded in parallel with speech. We demonstrate separating perception component without causing performance degradation of the speech recognition systems. We further demonstrate predicting EEG signals recorded in parallel with speaking from EEG signals recorded in parallel with passive listening and vice versa with very low normalized root mean squared error (RMSE) using Data set B used by authors in \cite{krishna2019state}. We finally demonstrate both isolated and continuous speech recognition using EEG signals recorded in parallel with listening, speaking, concatenation of listening, speaking  and improve the previous connectionist temporal classification (CTC) model \cite{graves2014towards} results demonstrated by authors in \cite{krishna2019state} using their data set B. In \cite{krishna2019state} authors didn't provide results for isolated speech recognition. Even though authors demonstrated EEG based speech recognition results using attention model \cite{chorowski2015attention} in \cite{krishna20,krishna2019state} and showed lower test time word error rates (WER) for smaller corpus size, the visualization of attention weights provided by authors in \cite{krishna20} demonstrate that attention model might be memorizing and not actually learning the underlying distribution when trained using smaller data sets. Further in \cite{krishna2019state} authors demonstrated that recurrent neural network (RNN) transducer model \cite{graves2013speech} demonstrated higher test time WER's, hence we only use CTC model for performing continuous speech recognition experiments in this paper.  We also demonstrate that EEG based isolated speech recognition results can be improved using siamese network \cite{koch2015siamese} when the training data set has only few samples per label.

\section{Regression Model}

Since EEG signals recorded in parallel with speaking and EEG signals recorded in parallel with passive listening for the same English sentence for the same subject was of different lengths we used encoder-decoder regression model with attention mechanism to predict EEG signals recorded in parallel with speaking from EEG signals recorded in parallel with passive listening and vice versa. Our encoder was a single layer of gated recurrent unit (GRU) \cite{chung2014empirical} with 128 hidden units and our decoder was also a GRU with 128 hidden units followed by a time distributed dense layer with 30 hidden units. The time distributed dense layer consists of linear activation function. The encoder GRU layer takes EEG features of dimension 30 as input and the encoder outputs are passed to luong dot product attention layer \cite{luong2015effective} to derive the attention context vectors which are passed to the decoder to get the predictions. The details of dot product attention calculations are explained in \cite{luong2015effective}. 
A dropout regularization \cite{srivastava2014dropout} with dropout rate 0.2 is applied after the attention layer. When listen EEG of dimension 30 is taken as input the model outputs spoken EEG of dimension 30 as output and vice versa. The details of the data set (Spoken, Listen EEG) are covered in the sections below. Basically the subjects were first asked to listen to English utterances and then they were instructed to speak out loud the utterances that they listened to. The term listen EEG refers to EEG signals recorded during passive learning and term spoken EEG refers to EEG signals recorded in parallel with speaking process. There was no fixed time step value for the input. 

The model was trained for 150 epochs using adam \cite{kingma2014adam} optimizer with mean squared error (MSE) as the loss function. The batch size was set to 100 and the validation split hyper parameter was set to a value of 0.1. The script was written using Keras deep learning framework. 

\section{Separation Model}

The goal of developing this deep model was to separate out speech perception component present in spoken EEG as one listen or hears his or her own speech during speaking process and thus the EEG signals recorded in parallel with speaking process or spoken EEG contains brain activity responsible for speech production as well as speech perception. Using EEG preprocessing methods discussed in the below sections it is possible to remove other biological signal artifacts from EEG signals to some extend but removing speech perception component using traditional signal processing methods is extremely challenging mainly due to the source localization issue and poor spatial resolution associated with EEG signals. The architecture of the separation model is described in Figure 1. The main idea here is to train a regression model and ASR model simultaneously in such a way that the ASR model helps in separating out the perception component but at the same time that it doesn't remove useful speech production components during the separation process. Our hypothesis is that the desired EEG signal or features can be modelled as a non-linear function of difference between spoken EEG and listen EEG features. The regression and ASR model are trained simultaneously to efficiently realize this non-linear function (tanh in our case). 
The model takes spoken EEG of dimension 30 as input as shown in Figure 1 and it passes through the regression part of the separation model. The regression part is similar to the regression model described before in the previous section but here we didn't use attention layer since the separation model needs to perform subtraction operation and hence the time steps value need to be preserved. Hence we instead performed trimming of Spoken EEG time steps to make it equal to Listen EEG time steps value. On an average we trimmed Spoken EEG time steps by 0.72 seconds. And we used temporal convolutional network (TCN) \cite{bai2018empirical} instead of GRU for faster training as the separation model is more complex than regression model. Since we didn't use attention layer, the decoder in the regression part consists of only the time distributed dense layer. 
The targets for the regression part is set to listen EEG features of dimension 30. The intermediate features outputted by the time distributed dense layer in the regression part is passed to a subtraction arithmetic block as shown in Figure 1. The subtraction arithmetic block calculates the difference between input spoken EEG features and the intermediate features. The difference features are then passed to tanh non-linearity through a fully connected dense layer and finally to a ASR classifier model. The architecture of the ASR classifier model is described in Figure 2. We used MSE as the loss function for regression part and cross entropy was used as the loss function for the ASR classifier model. The separation model was trained for 200 epochs using adam optimizer with a batch size of 50 until the combined MSE, cross entropy loss shows convergence and the validation split hyper parameter was set to a value of 0.1. During test time, the TCN layer in the separation model takes spoken EEG features of 30 as the input and we take output from the dense layer with tanh non-linearity ( the layer after the subtraction block).

\section{Isolated Speech recognition Model}

For performing isolated speech recognition experiments we used the same ASR classifier model described in Figure 2. The dense layer in the model consists of linear activation function. The last time step output of the  TCN layer is passed to the dense layer with two hidden units. The model was trained for 100 epochs using adam optimizer with a batch size of one. We used categorical cross entropy as the loss function for the model. The validation split hyper parameter was set to a value of 0.1. 

We also tried performing sentence identification task from a pair of given sentences using EEG features using the siamese network described by authors in \cite{koch2015siamese}. In \cite{koch2015siamese} authors used it for image recognition task. The siamese network is especially useful when we have few examples per label to train the classifier or deep model. 
Our siamese network consists of two TCN layers with 128 filters connected in parallel, the last time step output of each of the TCN layer is connected to a dense layer with 64 hidden units with sigmoid activation function to derive the embeddings for the given pair of the input. Then L1 distance is calculated between the embeddings. The L1 distance is passed to a dense layer with sigmoid activation. The model is trained for 100 epochs with batch size one and using adam as the optimizer. We used binary cross entropy as the loss function and the validation split hyper parameter was set to a value of 0.1. During test time a pair of EEG features are fed to the siamese network as input and the model outputs one if both features represents the same English sentence else, the model outputs zero. 

\section{Continuous Speech recognition Model}

For performing continuous speech recognition experiments we used the connectionist temporal classification (CTC) \cite{graves2006connectionist,graves2014towards} model described in Figure 1 in \cite{krishna2019improving} with the exact same hyper parameters and training parameters used by authors in \cite{krishna2019improving} but the encoder layers in the CTC model were initialized with random weights \cite{krishna20,krishna2019state}. An external language model was also used during inference time like the ones used by authors in \cite{krishna2019improving}. The CTC model we used in this work is different from the ones used by authors in \cite{krishna2019state}.


\begin{figure}[h]
\begin{center}
\includegraphics[height=5cm, width=1\linewidth,trim={0.1cm 0.1cm 0.1cm 0.1cm}]{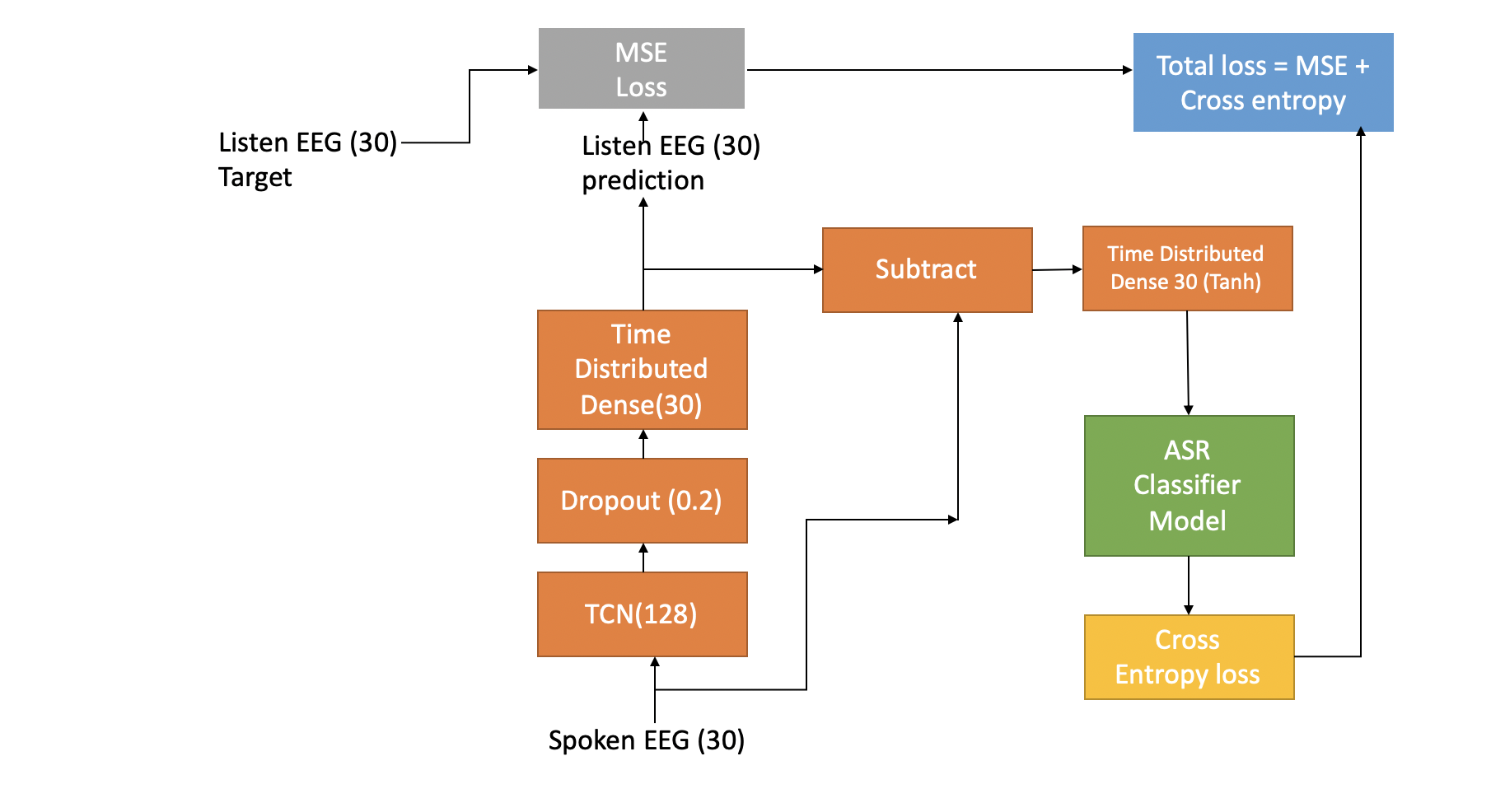}
\caption{Architecture of our proposed separation model} 
\label{1vsall}
\end{center}
\end{figure}

\begin{figure}[h]
\begin{center}
\includegraphics[height=4cm, width=\linewidth,trim={0.1cm 0.1cm 0.1cm 0.1cm}]{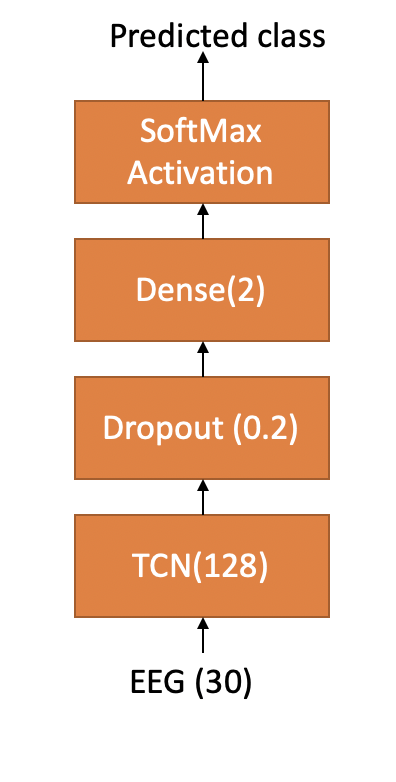}
\caption{Architecture of the ASR classifier model used in our proposed separation model} 
\label{1vsall}
\end{center}
\end{figure}




\section{Data Sets used for performing experiments}

For performing continuous speech recognition experiments using EEG we used Data set B used by authors in \cite{krishna2019state}. It basically consists of listen EEG and spoken EEG recordings from 15 subjects recorded in presence of a background noise of 50dB. 

For performing isolated speech recognition experiments we used first two unique sentences from Data set B used by authors in \cite{krishna2019state}, consisting of a total of 90 EEG recording examples for spoken, listen condition. The same data set was used to train our separation model described in Figure 1. Since there were only two unique sentences, hence the ASR classifier model's final dense layer had two hidden units with softmax activation function. We considered EEG samples for only two unique sentences since we were interested in faster training of the simultaneous regression and ASR classifier models. 
More details of the data set B, EEG experiment design, EEG recording hardware etc are covered in \cite{krishna2019state}. 

For training separation and regression model we used 80\% of the total data as training set, remaining 10\% as validation set and rest 10\% as test set. The train-test split was done randomly. There was no overlap between training, testing and validation set.
For training continuous speech recognition and isolated speech recognition models we performed experiments first where we used 80\% of the data as training set, remaining 10\% as validation set and rest 10\% as test set and then we performed experiments where we used data from first 13 subjects as training set, 14th subject data as validation set and last subject data as test set.

\section{EEG feature extraction details}
We followed the same EEG preprocessing methods used by authors in \cite{krishna2019speech,krishna20} for extracting raw EEG features for both spoken and listen conditions. 
The EEG signals were sampled at 1000Hz and a fourth order IIR band pass filter with cut off frequencies 0.1Hz and 70Hz was applied. A notch filter with cut off frequency 60 Hz was used to remove the power line noise.
The EEGlab's \cite{delorme2004eeglab} Independent component analysis (ICA) toolbox was used to remove other biological signal artifacts like electrocardiography (ECG), electromyography (EMG), electrooculography (EOG) etc from the EEG signals. 
We extracted five statistical features for EEG, namely root mean square, zero crossing rate,moving window average,kurtosis and power spectral entropy \cite{krishna2019speech,krishna20}. So in total we extracted 31(channels) X 5 or 155 features for EEG signals. The EEG features were extracted at a sampling frequency of 100Hz for each EEG channel.


\section{EEG Feature Dimension Reduction Algorithm Details}
After extracting EEG features as explained in the previous section, we used Kernel Principle Component Analysis (KPCA) \cite{mika1999kernel} to perform denoising of the EEG feature space as explained by authors in \cite{krishna20,krishna2019speech}. 
We reduced the 155 EEG features to a dimension of 30 by applying KPCA. We plotted cumulative explained variance versus number of components to identify the right feature dimension. We used KPCA with polynomial kernel of degree 3 \cite{krishna2019speech,krishna20}.

\section{Results}

We used word error rate (WER) as the performance metric to evaluate CTC model during test time, classification test accuracy was used as the performance metric to evaluate ASR classifier model or isolated speech recognition model during test time. The classification test accuracy is defined as the ratio of number of correct predictions given by model to total number of predictions given by the model on test set. The lower the WER value better is the continuous ASR system test time performance. 
For the regression model used for predicting EEG signals recorded in parallel with speaking (spoken EEG) from EEG signals recorded in parallel with passive listening (listen EEG) and vice versa we used normalized root mean squared error (RMSE) as the performance metric. The RMSE values were normalized by dividing the RMSE values with the absolute difference between the maximum and minimum value in the test set observation vector. The Tables 1,2 and 3 shows test time results obtained for continuous speech recognition where we used 10 \% of the total data set as test set (the same data split method used by authors in \cite{krishna2019state} to obtain results for their CTC model) and Table 4 shows test time results obtained for continuous speech recognition where we used last subject data as test set. Tables 5 and 6 shows the test time results obtained for isolated speech recognition using ASR classifier model. For spoken+listen condition, we concatenate spoken and listen EEG features along time step axis for each English sentence. 
We observed an average improvement of 5 \% in test accuracy for spoken, listen, spoken + listen conditions using siamese network over ASR classifier model for the sentence identification task from a pair of input EEG features. 


The overall results from Tables 3,4,5 and 6 shows that speech perception components present in EEG provide useful features to the ASR model as the test time results for spoken + listen EEG condition was better than spoken EEG condition for majority of the ASR experiments. However when we design a reliable EEG based speech prosthetic it should only use EEG features responsible for speech production. In Table 1 we show the test time results obtained using spoken EEG features with CTC model after removing perception components using our separation model. We can observe that our separation model was able to remove perception components from spoken EEG without causing much performance degradation for performing recognition. In Tables 1 and 2 the results under Ref[1] column were directly taken from the results mentioned by authors in \cite{krishna2019state} for CTC model for Data set B under their 'Results' portion mentioned just before the beginning of their Table 1. In \cite{krishna2019state} authors didn't provide results for isolated speech recognition for listen or spoken or listen + spoken condition.   

For predicting spoken EEG features from listen EEG features using the regression model we observed a normalized RMSE of \textbf{ 0.0016446532} and for predicting listen EEG features from spoken EEG features we observed a normalized RMSE of \textbf{0.0052599716} during test time.

\begin{table}[!ht]
\centering
\begin{tabular}{|l|l|l|l|l|}
\hline
\textbf{\begin{tabular}[c]{@{}l@{}}Total\\ Number\\ of\\ Sentences\end{tabular}} & \textbf{\begin{tabular}[c]{@{}l@{}}Total\\ No\\ of\\ words\end{tabular}} & \textbf{\begin{tabular}[c]{@{}l@{}}Spoken\\ EEG\\ WER\\ (\%)\\ REF\\ {[}1{]}\end{tabular}} & \textbf{\begin{tabular}[c]{@{}l@{}}Spoken\\ EEG\\ WER\\ (\%)\\ our\\ CTC\\ Model\end{tabular}} & \textbf{\begin{tabular}[c]{@{}l@{}}Spoken\\ EEG\\ WER\\ (\%)\\ after\\ removing\\ perception\\ component\end{tabular}} \\ \hline
27                                                                               & 173                                                                      & 73.6                                                                                       & 66.6                                                                                           & \textbf{62.7}                                                                                                          \\ \hline
45                                                                               & 292                                                                      & 83.8                                                                                       & \textbf{76.53}                                                                                 & 77.12                                                                                                                  \\ \hline
63                                                                               & 404                                                                      & 91.1                                                                                       & 81                                                                                             & \textbf{79.18}                                                                                                         \\ \hline
81                                                                               & 525                                                                      & 91.5                                                                                       & 84.2                                                                                           & \textbf{82.72}                                                                                                         \\ \hline
\end{tabular}
\caption{CTC Model continuous speech recognition test time results using spoken EEG where total data was randomly splitted to form the test set. Number of unique sentences and unique words contained in test set were same as the ones mentioned in Table 3 in \cite{krishna2019state}}
\end{table}

\begin{table}[!ht]
\centering
\begin{tabular}{|l|l|l|l|}
\hline
\textbf{\begin{tabular}[c]{@{}l@{}}Total\\ Number\\ of\\ Sentences\end{tabular}} & \textbf{\begin{tabular}[c]{@{}l@{}}Total\\ No\\ of\\ words\end{tabular}} & \textbf{\begin{tabular}[c]{@{}l@{}}Listen\\ EEG\\ WER\\ (\%)\\ REF\\ {[}1{]}\end{tabular}} & \textbf{\begin{tabular}[c]{@{}l@{}}Listen\\ EEG\\ WER\\ (\%)\\ our\\ CTC\\ Model\end{tabular}} \\ \hline
27                                                                               & 173                                                                      & \textbf{52.6}                                                                              & 75.24                                                                                          \\ \hline
45                                                                               & 292                                                                      & 87.09                                                                                      & \textbf{74.83}                                                                                 \\ \hline
63                                                                               & 404                                                                      & 88.88                                                                                      & \textbf{73.71}                                                                                 \\ \hline
81                                                                               & 525                                                                      & 94.9                                                                                       & \textbf{77.17}                                                                                 \\ \hline
\end{tabular}
\caption{CTC Model continuous speech recognition test time results using listen EEG where total data was randomly splitted to form the test set}
\end{table}

\begin{table}[!ht]
\centering
\begin{tabular}{|l|l|l|}
\hline
\textbf{\begin{tabular}[c]{@{}l@{}}Total\\ Number\\ of\\ Sentences\end{tabular}} & \textbf{\begin{tabular}[c]{@{}l@{}}Total\\ No\\ of\\ words\end{tabular}} & \multicolumn{1}{c|}{\textbf{\begin{tabular}[c]{@{}c@{}}Listen\\ +\\ spoken\\ EEG\\ WER\\ (\%)\\ our\\ CTC\\ Model\end{tabular}}} \\ \hline
27                                                                               & 173                                                                      & 56.56                                                                                                                            \\ \hline
45                                                                               & 292                                                                      & 79                                                                                                                               \\ \hline
63                                                                               & 404                                                                      & 80.7                                                                                                                             \\ \hline
81                                                                               & 525                                                                      & 84                                                                                                                               \\ \hline
\end{tabular}
\caption{CTC Model continuous speech recognition test time results using concatenation of listen and spoken EEG where total data was randomly splitted to form the test set}
\end{table}

\begin{table}[!ht]
\centering
\begin{tabular}{|l|l|l|l|}
\hline
\textbf{\begin{tabular}[c]{@{}l@{}}Total\\ Number\\ of\\ Sentences\end{tabular}} & \textbf{\begin{tabular}[c]{@{}l@{}}Spoken\\ EEG\\ WER\\ (\%)\\ our\\ CTC\\ Model\end{tabular}} & \multicolumn{1}{c|}{\textbf{\begin{tabular}[c]{@{}c@{}}Listen\\ EEG\\ WER\\ (\%)\\ our\\ CTC\\ Model\end{tabular}}} & \textbf{\begin{tabular}[c]{@{}l@{}}Listen\\ +\\ Spoken\\ EEG\\ WER\\ (\%)\\ our\\ CTC\\ Model\end{tabular}} \\ \hline
27                                                                               & 81.9                                                                                           & 71.4                                                                                                                & 78.78                                                                                                       \\ \hline
\end{tabular}
\caption{CTC Model continuous speech recognition test time results where last subject data was used as test set. Number of unique sentences was 9 and unique words contained in test set was 55}
\end{table}

\begin{table}[!ht]
\centering
\begin{tabular}{|l|l|l|}
\hline
\textbf{\begin{tabular}[c]{@{}l@{}}Spoken\\ EEG\\ \% Test\\ acc\end{tabular}} & \textbf{\begin{tabular}[c]{@{}l@{}}Listen\\ EEG\\ \% Test\\ acc\end{tabular}} & \multicolumn{1}{c|}{\textbf{\begin{tabular}[c]{@{}c@{}}Listen\\ EEG\\ +\\ Spoken\\ EEG\\ \% Test\\ acc\end{tabular}}} \\ \hline
50                                                                            & 50                                                                            & 55.56                                                                                                                 \\ \hline
\end{tabular}
\caption{Isolated speech recognition test time results using ASR classifier model where total data was randomly splitted to form the test set}
\end{table}

\begin{table}[!ht]
\centering
\begin{tabular}{|l|l|l|}
\hline
\textbf{\begin{tabular}[c]{@{}l@{}}Spoken\\ EEG\\ \% Test\\ acc\end{tabular}} & \textbf{\begin{tabular}[c]{@{}l@{}}Listen\\ EEG\\ \% Test\\ acc\end{tabular}} & \multicolumn{1}{c|}{\textbf{\begin{tabular}[c]{@{}c@{}}Listen\\ EEG\\ +\\ Spoken\\ EEG\\ \% Test\\ acc\end{tabular}}} \\ \hline
50                                                                            & 50                                                                            & 66.67                                                                                                                 \\ \hline
\end{tabular}
\caption{Isolated speech recognition test time results using ASR classifier model where last subject data was used as test set}
\end{table}

\section{Conclusion and Future work}
In this paper we introduced a deep model to separate speech perception component from EEG signals recorded in parallel with speech. We further demonstrated isolated and continuous speech recognition using EEG features recorded under various conditions (listen, spoken). We finally demonstrated predicting EEG signals recorded in parallel with speaking (spoken EEG) from EEG signals recorded in parallel with passive listening (listen EEG) and vice versa with very low normalized root mean squared error (RMSE) during test time. Future work will focus on validating and improving the results using a larger data set and also perform separation experiments by replacing the ASR classifier model with CTC model. Training a CTC model instead of a simple ASR classifier model within the separation model would require a larger data set.

\section{Acknowledgement} 
We would like to thank Kerry Loader and Rezwanul Kabir from Dell, Austin, TX for donating us the GPU to train the models.   

\bibliographystyle{IEEEtran}

\bibliography{mybib}


\end{document}